\documentclass[aps,prx, amsmath,amssymb,twocolumn,superscriptaddress]{revtex4-2}

\usepackage{graphicx}
\usepackage{bm}
\usepackage{bbold}
\usepackage{hyperref}
\usepackage{dsfont}
\usepackage{color}

\usepackage{ulem}
\usepackage{xcolor}
\usepackage[utf8]{inputenc} 
\usepackage[T1]{fontenc}
\usepackage{xspace}

\renewcommand{\i}{\boldsymbol{i}}

\newcommand{\ve}[1]{\boldsymbol{#1}}

\begin{document}

\title{ Tuning the order of a deconfined quantum critical point}
\author{Anika  G\"otz}
\affiliation{\mbox{Institut f\"ur Theoretische Physik und Astrophysik,
		Universit\"at W\"urzburg, 97074 W\"urzburg, Germany}}
\author{Fakher F. Assaad}
\affiliation{\mbox{Institut f\"ur Theoretische Physik und Astrophysik,
    Universit\"at W\"urzburg, 97074 W\"urzburg, Germany}}
\affiliation{W\"urzburg-Dresden Cluster of Excellence ct.qmat, Am Hubland, 97074 W\"urzburg, Germany}
\author{Natanael C. Costa }
\affiliation{Instituto de Fisica, Universidade Federal do Rio de Janeiro Cx.P. 68.528, 21941-972 Rio de Janeiro RJ, Brazil}

%\date{\today}

\begin{abstract}
  We  consider a Su-Schrieffer-Heeger model in  the  assisted hopping limit, where    direct  electron hopping is subdominant.  At fixed  electron-phonon coupling and in the 
  absence of  Coulomb  interactions,  the model shows  a  deconfined quantum critical point (DQCP)  between a  $(\pi,0)$  valence bond  solid  in the  adiabatic limit   and  a 
  quantum antiferromagnetic (AFM) phase at  high phonon frequencies.  Here,  we  show  that by 
  adding terms to the model that reinforce the AFM  phase,  thereby  lowering  the  
  critical phonon  frequency, the  quantum phase transition  becomes  strongly  first 
 order.  Our  results   do  not  depend  on the  symmetry of  the model. In fact,  adding  a  Hubbard-$U$  term to the model   lowers  the O(4) symmetry of  the model  to 
 SU(2)   such  that  the   DQCP we observe  has  the  same  symmetries  as  other 
 models that   account  for  similar quantum phase  transitions.   
\end{abstract}

\maketitle
\textit{Introduction.} Deconfined quantum criticality  (DQC) \cite{Senthil04_1,Senthil04_2,WangC17} refers to a  direct and continuous quantum phase transition between two different  broken symmetry states. This lies 
at odds with order parameter based Ginzburg–Landau–Wilson  (GLW)  approaches that would  generically  predict a first-order transition.  The missing element is topology.  A very natural 
way to  understand  DQC  is in terms of Dirac  fermions  and compatible or  intertwined mass terms \cite{Tanaka05,Abanov00,Senthil06,Ryu09}  that describe various orders. For  example, in  
1+1 dimensions, compatible mass terms include the three  antiferromagnetic (AFM)  and single   valence bond solid  (VBS) masses.  The algebraic requirement that compatible 
mass terms anticommute leads to the result that topological  defects of one order carry the charge of the other order. For  example, in  the 1+1D setting,  a domain wall  of the VBS order 
hosts a spin-1/2  degree of  freedom.  In 2+1D, the three  AFM  and  two VBS masses lead to the very  same effect, namely that 
a  VBS vortex carries a spin-1/2  degree of freedom \cite{Levin04}.  In 1+1D, the  critical point is know to be captured  by  the SO(4)   non-linear sigma model with Wess-Zumino-Witten term. 
At the  critical point,  we  observe an enhanced  symmetry that allows to rotate   between VBS and AFM orders.    The theory has a marginal operator that breaks  SO(4) symmetry  down to  $\mathrm{SO}(3) \times \mathbb{Z}_2$: it is marginally relevant (irrelevant)  on the  VBS (AFM) side.   We note  that other  instances of DQC  have been observed  in models  where  fermions are not gapped out \cite{LiuZH21,LiuZ24} and  that do not fit in the aforementioned picture of compatible  Dirac mass terms.  

Here, we will concentrate on the  2+1-dimensional  bosonic  case.   Interestingly, there are many models, spin models \cite{Sandvik07},  loop models \cite{Nahum15}, fermion models \cite{Liu18}  and  electron-phonon models \cite{Goetz23},  which at finite energy scale all point to the 
same phenomenology   as in 1+1 dimensions but with emergent SO(5)  symmetry \cite{Nahum15_1,SatoT22}.  However,   the numerical value of the correlation length exponent lies at odds  with conformal bootstrap bounds  required to guarantee a single relevant operator \cite{Nakayama16}. In other words,  the  very existence of a critical point.    This, as well as observed violations to scaling \cite{Nahum15} at large distances, poses  a lot of questions.  Various  scenarios including complex fixed points \cite{WangC17,WangC19,Nahum19} as well as the possibility of a tricritical point have been put  forward \cite{Chester24,Takahashi24}.    Both scenarios imply that there must be a tuning parameter  that does not break the  underlying symmetries but  that renders the deconfined quantum critical point (DQCP) strongly first order.  

\begin{figure}[t]
  \includegraphics[width=0.8\linewidth]{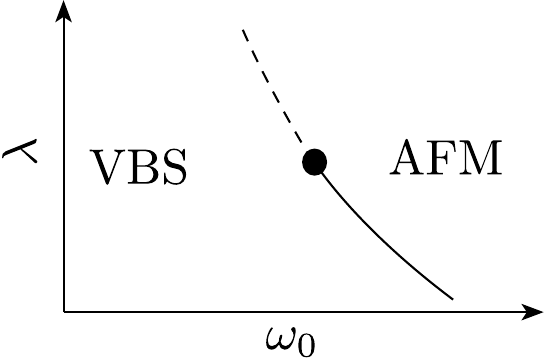}
   \caption{\label{fig:sketch} The solid (dashed) line  corresponds to a continuous (first order) transition. The model parameter $\lambda$ reduces  the value of the critical phonon frequency, $\omega_0$, at which we observe  the VBS to AFM transition.  }
 \end{figure}

We consider  a novel realization of  DQC  in a  Su-Schrieffer-Heeger  \cite{Su80} model  defined on a  square lattice \cite{Goetz23}.  The  key point is  that the tuning parameter  to observe  DQC  is  the  phonon frequency. By changing  other parameters in the model, we can tune the 
critical phonon frequency.   We  observe  that this  allows  to tune  the DQCP  from a continuous or 
weakly first order  transition to a strong first order one (see Fig.~\ref{fig:sketch}).  The interpretation of  our result is not unique: the bullet point in Fig.~\ref{fig:sketch} can be interpreted in terms of 
a Peierls transition of the U(1) gauge theory realized at the DQCP \cite{Seifert23,Hofmeier24}, in terms of 
a complex SO(5) CFT stemming from fix-point annihilation \cite{Nahum19,WangC19,WangZ20}, or in terms of an 
SO(5) multi-critical point \cite{Chester24,Takahashi24}. 

\textit{Model, symmetries and  method.}
The model   we  consider  has  fermion degrees  of  freedom  on the  sites, $\bm{i}$, of a square lattice and 
bosonic  degrees  of  freedom  on the bonds, $b$:
\begin{eqnarray}
\label{Eq:model}
   \hat{H}  & & =  \sum_{b} \left[ (-t +  g\hat{Q}_{b})  \hat{K}_b  - \lambda \hat{K}_b^{2}  +  
   \frac{\hat{P}_b^2}{2M}  +   \frac{1}{2}  \hat{Q}_{b}^2 \right]  +  \nonumber  \\
    & &  +  \frac{U}{2} \sum_{\bm{i}}\left( \hat{n}_{\bm{i}}  -1 \right)^2\,.
\end{eqnarray}
The hopping operator is defined as  $ \hat{K}_{b=\langle \bm{i},\bm{j}\rangle } =  \sum_{\sigma=1}^{2}( \hat{c}^{\dagger}_{\bm{i},\sigma} 
\hat{c}^{\phantom\dagger}_{\bm{j},\sigma}  + \hat{c}^{\dagger}_{\bm{j},\sigma} \hat{c}^{\phantom\dagger}_{\bm{i},\sigma} )  $   with the fermion operators $\hat{c}^{\phantom\dagger}_{\bm{i},\sigma} $, $\hat{n}_{\bm{i}}  =  \sum_{\sigma=1}^{2} 
\hat{c}^{\dagger}_{\bm{i},\sigma}  \hat{c}^{\phantom\dagger}_{\bm{i},\sigma}  $, while 
$\hat{P}_b$ and  $\hat{Q}_b $ are the   momentum  and   position   operators of a   harmonic  oscillator.  With a canonical transformation of the bosonic modes, we can set the spring constant to unity, thus justifying  the   $\frac{1}{2}  \hat{Q}_{b}^2$ term.  In the notation of Eq.~(\ref{Eq:model}), the model is defined by  the hopping $t$,  the  phonon frequency, $\omega_0  =  \sqrt{\frac{1}{m}} $,  by  the electron-phonon coupling $\lambda_{e-ph} =  \frac{g^2}{2} $  as  well as  the magnitude of the  square hopping $\lambda$.  In the runs presented below, we  set the unit by fixing   $\lambda_{e-ph} = 2 $. We consider $t=0.1$  and vary $\lambda$, $U$ and the phonon frequency.     At  $U=0$,  this  model  has  been studied in details in   Ref.~\cite{Goetz23}.  

A  key  point of the model in the small  $t$ limit  is the emergence  of  a  $\pi$  flux per plaquette  that  results  in  emergent Dirac  fermions \cite{Goetz23}.  As discussed  in the  introduction, this sets the stage for exotic quantum criticality. 
The  value  of  $t$ is  such  that the model  harbors a  $\pi$ flux per plaquette.   At  moderate  values of  $\lambda = 0.5$ considered  in  Ref.~\cite{Goetz23}   and  at  $U=0$,  we  observe  a direct  
and  continuous  transition between a $(\pi,0) $ VBS  and   AFM  phase   as  a  function  of  increasing  phonon  frequency, $ \omega_0 $.  The aim of  this article is  twofold.  First,  we  will   augment  the value of  $\lambda$ in the  O(4)  model.   Since    for  
$b=\langle \bm{i},\bm{j}\rangle$
\begin{equation}
     -\frac{1}{4} \hat{K}_b^2   =  \hat{\ve{S}}_{\bm{i}} \cdot  \hat{\ve{S}}_{\bm{j}}   +  \hat{\ve{\eta}}_{\bm{i}}  \cdot \hat{\ve{\eta}}_{\bm{j}} \,, 
\end{equation}
larger  values of  $\lambda$   will   favor antiferromagnetism  and  thereby  lower  the   critical frequency  at  which  the  transition of  VBS  to AFM occurs.  Here,  $\hat{\ve{S}}_{\bm{i}}  =    \frac{1}{2}  \hat{\ve{c}}^{\dagger}_{\bm{i}}   \ve{\sigma} \hat{\ve{c}}^{\phantom\dagger}_{\bm{i}}  $  and  the  Anderson  pseudospin  operator $\hat{\ve{\eta}}_{\bm{i}} =  \hat{P}^{-1}\hat{\ve{S}}_{\bm{i}} \hat{P} $  with   $\hat{P}$   the partial particle-hole symmetry: 
\begin{equation}
  \hat{P}^{-1} \hat{c}^{\dagger}_{\bm{i},\sigma} \hat{P}   = \delta_{\sigma,\uparrow}   \hat{c}^{\dagger}_{\bm{i},\sigma}   + \delta_{\sigma,\downarrow} e^{i \ve{Q}\cdot \ve{i}}  \hat{c}^{\phantom\dagger}_{\bm{i},\sigma}\,,
\end{equation}
 with $ \ve{Q}  = (\pi,\pi) $,  in units where the lattice constant is set to unity.  Second,  we will  add  a  Hubbard-$U$ term so  as  to  reduce the O(4)  symmetry  to SO(4).

The O(4) symmetry of the model at $U=0$   becomes apparent when writing: 
\begin{equation}
    \hat{K}_{b=\langle \bm{i},\bm{j}\rangle} =   -\frac{i}{2}\sum_{n,\sigma} \hat{\gamma}_{\bm{i},n,\sigma} \hat{\gamma}_{\bm{j},n,\sigma} \,,
\end{equation}
where  $\hat{\gamma}_{\bm{i},n,\sigma}$ are Majorana fermions.  Here, $n$ labels the real and imaginary parts of the  fermion operator. Combining $n$ and $\sigma$ into a  four-component index, $\alpha$, explicitly  shows  the O(4)  symmetry:
\begin{equation} 
  \hat{\gamma}_{\bm{i},\alpha}   \rightarrow   \sum_{\alpha'=1}^{4} O_{\alpha,\alpha'} \hat{\gamma}_{\bm{i},\alpha'} 
\end{equation}
with  $O$ an  O(4) matrix. 
Since  partial particle-hole symmetry    leaves the Hamiltonian invariant,  the  AFM  state is degenerate with the $s$-wave  superconducting state and a charge density wave state.  

In the Majorana  representation,  the  Hubbard-$U$ term  reads: 
\begin{equation}
    H_U  =  U  \sum_{\bm{i}}  \hat{\gamma}_{\bm{i},1}  \hat{\gamma}_{\bm{i},2}   \hat{\gamma}_{\bm{i},3}   \hat{\gamma}_{\bm{i},4} \,.  
\end{equation}
Under an O(4)  rotation, the Hubbard interaction   transforms as
\begin{equation}
  H_U  \rightarrow  \det(O)U  \sum_{\bm{i}}  \hat{\gamma}_{\bm{i},1}  \hat{\gamma}_{\bm{i},2}   \hat{\gamma}_{\bm{i},3}   \hat{\gamma}_{\bm{i},4}   \,.
\end{equation}
As a  consequence, the  O(4)   symmetry  is reduced to  SO(4).

From  the  technical  point of view,   the  presence  of   the $\lambda$-term   allows  us  to adapt  a  method  proposed in  Ref.~\cite{Karakuzu18}  to  integrate out the fermions. As shown in  Ref.~\cite{Goetz23},   this approach turns out to be  efficient in terms  of  autocorrelation  times.   We have  used  the  Algorithms  for 
Lattice Fermions (ALF)  \cite{ALF_v1,ALF_v2}   implementation of  the auxiliary-field 
quantum Monte Carlo  algorithm  \cite{Blankenbecler81,White89,Assaad08_rev} to  carry  out  the numerical  simulations. The reader is  referred  to  \cite{Goetz23}  for a    detailed account  of our  implementation.  For  our  
simulations,  we have used a  symmetric Trotter  decomposition and set  the imaginary time step to $\Delta\tau=0.05$.

\textit{Results.} The  question we  will ask  is   if  the nature of  the  quantum  phase  transition changes   as  the critical phonon  frequency     diminishes.   We  will address the  very  same  question  for the O(4) ($U=0$)  and  SO(4) (finite $U$)  models  and we will see  that the very same results   hold. 

\begin{figure}[t]
 \includegraphics[width=\linewidth]{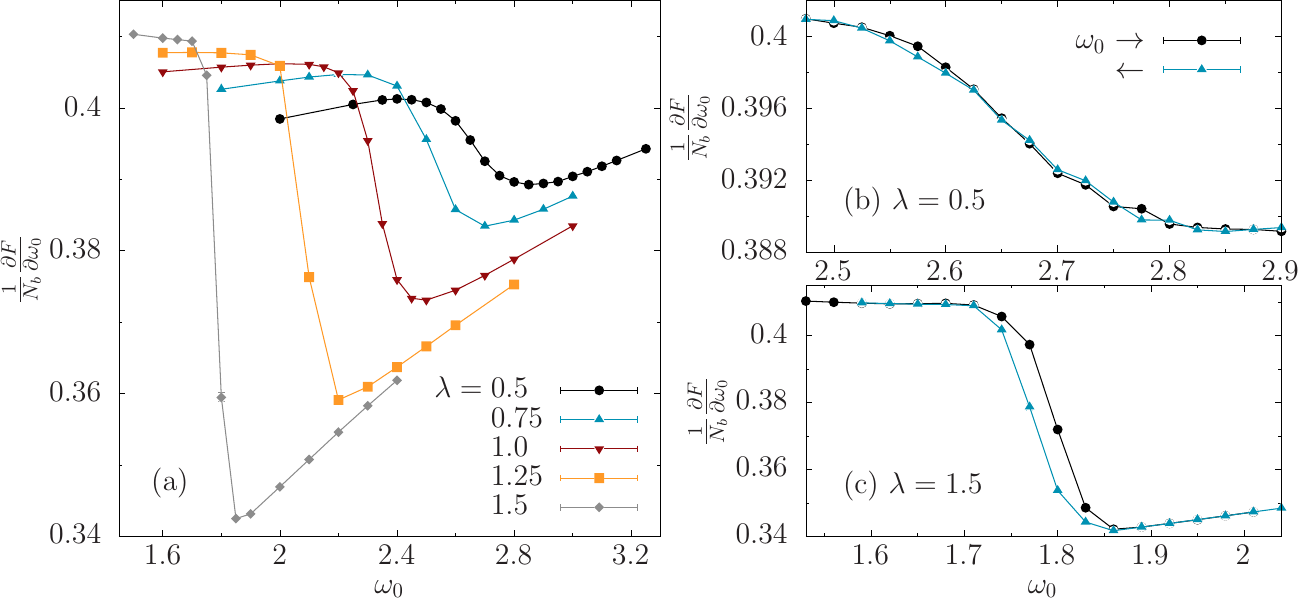}
  \caption{\label{fig:scal_hyst_O4}(a) Normalized free-energy derivative with
    respect to $\omega_0$. (b) and (c) Hysteresis curve for free-energy derivative with respect to 
  $\omega_0$. Here, $t=0.1$, $U=0$, and $\beta=L=8$. }
\end{figure}

\begin{figure}[b]
 \includegraphics[width=1\linewidth]{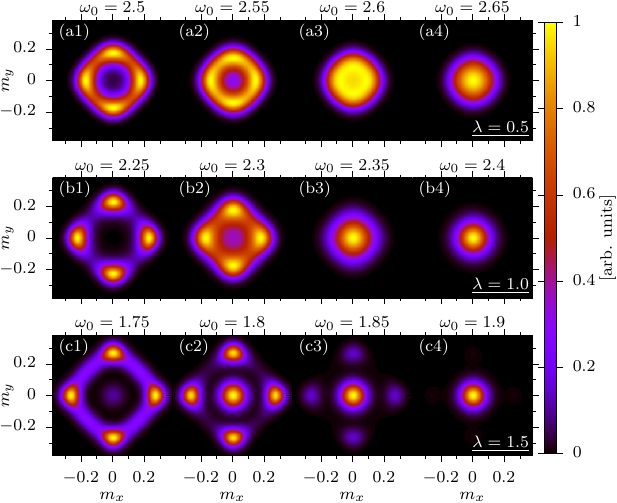}
  \caption{\label{fig:hist_O4}Histogram of the VBS order parameter 
    $m_x$ and $m_y$  for different $\lambda$ and $\omega_0$ at $t=0.1$, $U=0$ and
    $\beta=L=14$.}
\end{figure}

\textit{O(4).}
Figure~\ref{fig:scal_hyst_O4}(a)  plots the derivative of the free energy with respect to the phonon frequency
\begin{equation}
      \frac{\partial F}{\partial \omega_0}  = m \omega_0 \sum_b \langle \hat{Q}_b^2 \rangle + \frac{g}{\omega_0} \sum_b \langle \hat{Q}_b \hat{K}_b \rangle
\end{equation}
for various values of  $\lambda$.  As apparent,  as  $\lambda$  increases,  a  step-like feature   develops,  thereby signaling a first-order transition. The data at $\lambda = 0.5 $ correspond  to the data  taken in Ref.~\cite{Goetz23}, 
where we  argued in favor  of a  DQCP  between the AFM and VBS state at a critical phonon frequency  $\omega_0^{c} \simeq  2.6 $.   
The VBS  breaks the $C_4$ symmetry  and is a fourfold degenerate state.  To capture  this symmetry 
breaking, we  compute  the dimer-dimer correlation functions:
\begin{eqnarray}\label{eq:dimer_order}
  \hat{\Delta}_{\mu}(\bm{q})  &=& \frac{1}{\sqrt{N_s}} \sum_{\i} e^{i \bm{q}\cdot \i} \hat{\Delta}_{\i,\mu} \, , \nonumber\\
  \hat{\Delta}_{\i,\mu} & = & \hat{S}_{\sigma,\rho}(\i) \hat{S}_{\rho,\sigma}(\i+\bm{a}_{\mu}) \,, 
  \end{eqnarray}
  where 
  \begin{equation}
  \hat{S}_{\sigma,\rho}(\i) = \hat{c}^{\dagger}_{\i,\sigma} \hat{c}^{\phantom{\dagger}}_{\i,\rho} - \frac{1}{2} \delta_{\sigma,\rho} \,. 
  \end{equation}
 Figure~\ref{fig:hist_O4}  corresponds  to  histograms  of the order parameters $m_{\mu} = \hat{\Delta}_{\mu}(\bm{q}_{\mu})/\sqrt{N_s}$ with $\mu=x,y$,  $\bm{q}_{x}=(\pi,0)$,  $\bm{q}_{y}=(0,\pi)$ and $N_s$ the number of lattice sites \footnote{As in Ref.~\cite{Goetz23}, we symmetrized the histograms by exploiting the $C_4$ symmetry of the model and the arbitrariness of the minus sign in the definition of the order parameter in Eq.~(\ref{eq:dimer_order}).}.

   Deep  in the  VBS-ordered phases [Figs.~\ref{fig:hist_O4}(a1)(b1)(c1)], the histograms clearly  show  four peaks,  thereby demonstrating the fourfold degeneracy.   At $\lambda = 0.5$, where  we expect a DQCP,  the  $C_4$ symmetry gives  way  to an emergent U(1) symmetry that  shows up   in the form of a circular histogram in the vicinity of the critical point [Fig.~\ref{fig:hist_O4}(a3)]. In the AFM  phase, the  histogram shows a point-like feature  at $m_x = m_y = 0$ [Fig.~\ref{fig:hist_O4}(a4)].   In  contrast,  for a strongly first-order transition, one  expects  to observe  a  coexistence region of  VBS  and  AFM. This is  precisely seen in  Figs.~\ref{fig:hist_O4}(c2)(c3)  at  $\lambda = 1.5$, where both the four-peak structure and the central peak are simultaneously present.  The results from the  histograms are  confirmed by the 
hysteresis  curves   of Figs.~\ref{fig:scal_hyst_O4}(b) and \ref{fig:scal_hyst_O4}(c). Here, the final configuration of a simulation at $\omega_0$ is used as a starting configuration of a simulation at $\omega_0 \pm \Delta\omega_0$.

\textit{SO(4).} Models of  DQCP  possess an  $\mathrm{SU}(2) \times C_4$   \cite{Sandvik07,Nahum15}  or an $\mathrm{SU}(2) \times \mathrm{U}(1)$  symmetry \cite{Liu18}.    The SO(4)  symmetry  accounts  for  SU(2) spin symmetry  and  SU(2)  symmetry  of the  $\eta$ operators.  For  a  repulsive  Hubbard $U$,  even parity sites that define the Hilbert space on which the $\eta$ operators act  correspond to excited states. Hence, in the low  energy limit, our model   with the Hubbard-$U$ term has  the same symmetry as  the aforementioned lattice models of  DQCP.     Throughout this section,  we will consider  $U=0.5$,  and  retain the same value of  $t=0.1$.  We will again vary  $\lambda$ and  the phonon frequency. 

\begin{figure}[t]
  \includegraphics[width=1\linewidth]{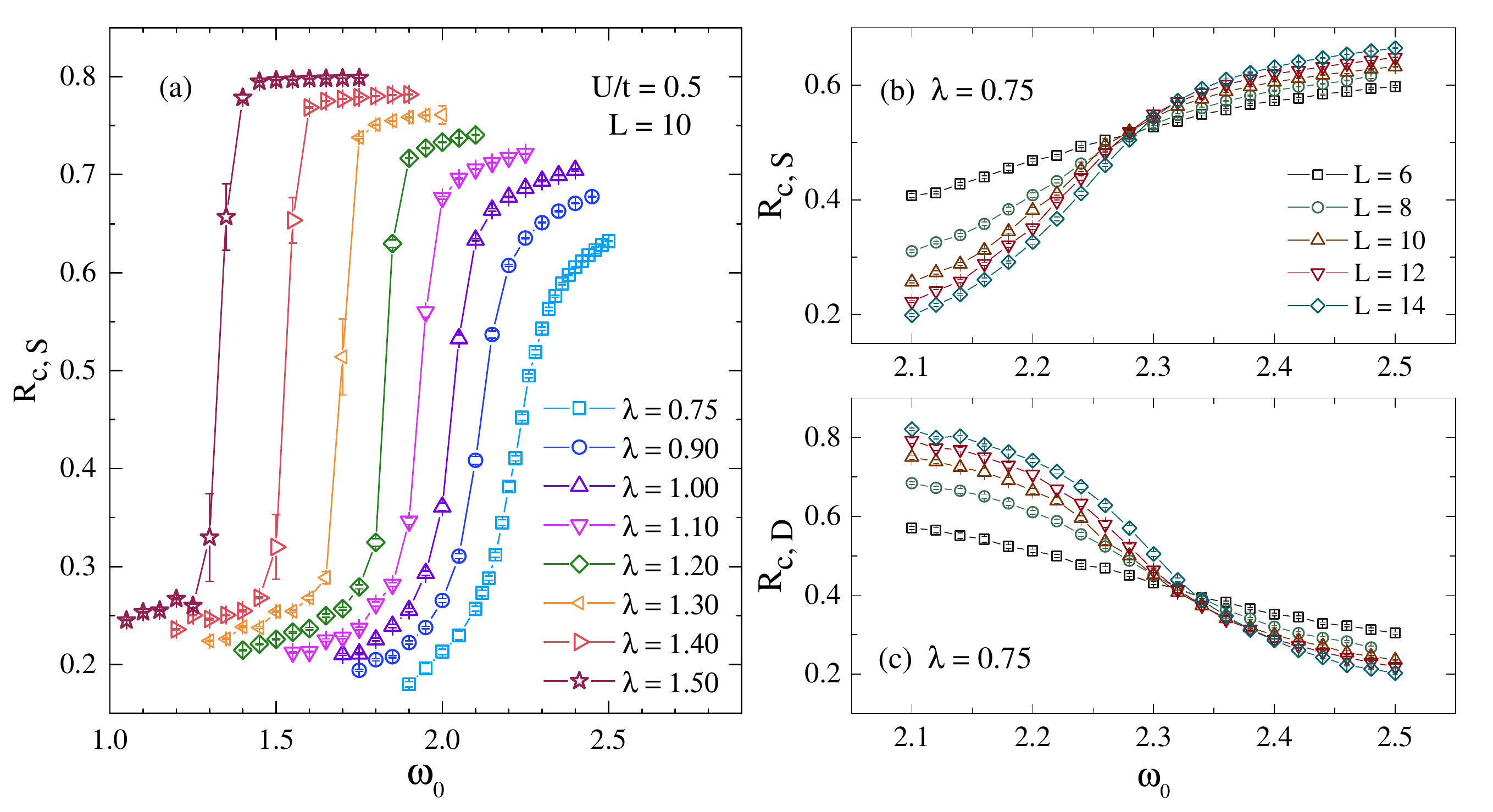} \caption{\label{fig:SSHH1} (a) Spin correlation ratio as a function of  $\lambda$ at $\beta = L = 10$. We see  that upon increasing $\lambda$  the correlation ratio develops a discontinuity.  (b) Spin correlation ratio  at $\lambda = 0.75$ and as a function of system size. (c)  Dimer correlation at $\lambda=0.75$  as a function of system size. For all the plots,  we fixed $U=0.5$, $\beta=L$, and $t=0.1$.}
 \end{figure}
The  first point to confirm  is  that at  small values of $\lambda$   the  numerical data   supports the  point of view  of a   DQCP, at least on our considered lattice sizes.
Here,  we will concentrate on the correlation ratio defined  as  
\begin{equation}
  R_{c,O}    =  1  -   \frac{S_O(\ve{Q}+\Delta\ve{q})}{S_O(\ve{Q})}\,,
\end{equation} 
where $  S_O(\ve{q}) =  \frac{1}{N_s} \sum_{\ve{i},\ve{j}} e^{i\ve{q} \cdot(\ve{i} - \ve{j}) } \langle \hat{O}(\ve{i}) \hat{O}(\ve{j}) \rangle  $ is a correlation function of a local observable $\hat{O}(\ve{i})$. We consider both the spin-spin correlations as  well as the dimer-dimer  correlations, see Eq.~(\ref{eq:dimer_order}).
In the  thermodynamic limit, $R_c$  converges to  unity (zero)   in the ordered (disordered) phase. $R_c$ is  a renormalization group  invariant quantity,  such  that in the vicinity of a critical point we expect $
R_{c,O}  \simeq   f( (\omega_0 -  \omega_0^c)L^{1/\nu}, L/\beta^{z}, L^{-\omega} )$.
Here, $\nu$ is the correlation length  exponent, $z$ the  dynamical exponent and 
$\omega$ the leading correction to the  scaling exponent.   As we will confirm below, at criticality we observe Lorentz symmetry such that we can set $z=1$ and adopt a $\beta = L$ scaling.  
In the  absence of corrections to scaling, the $R_c$  curves on various lattice sizes  all cross at  the critical point. Figures~\ref{fig:SSHH1}(b) and \ref{fig:SSHH1}(c) show the correlation ratio as a function of  system size for both the VBS and AFM correlations.  The  position of the crossing can only be approximately  determined.  However, within our accuracy,  the data is consistent with  a direct transition between the AFM and VBS phase  at $\omega_0^c \simeq 2.25-2.35$.  

\begin{figure}[t]
  \includegraphics[width=\linewidth]{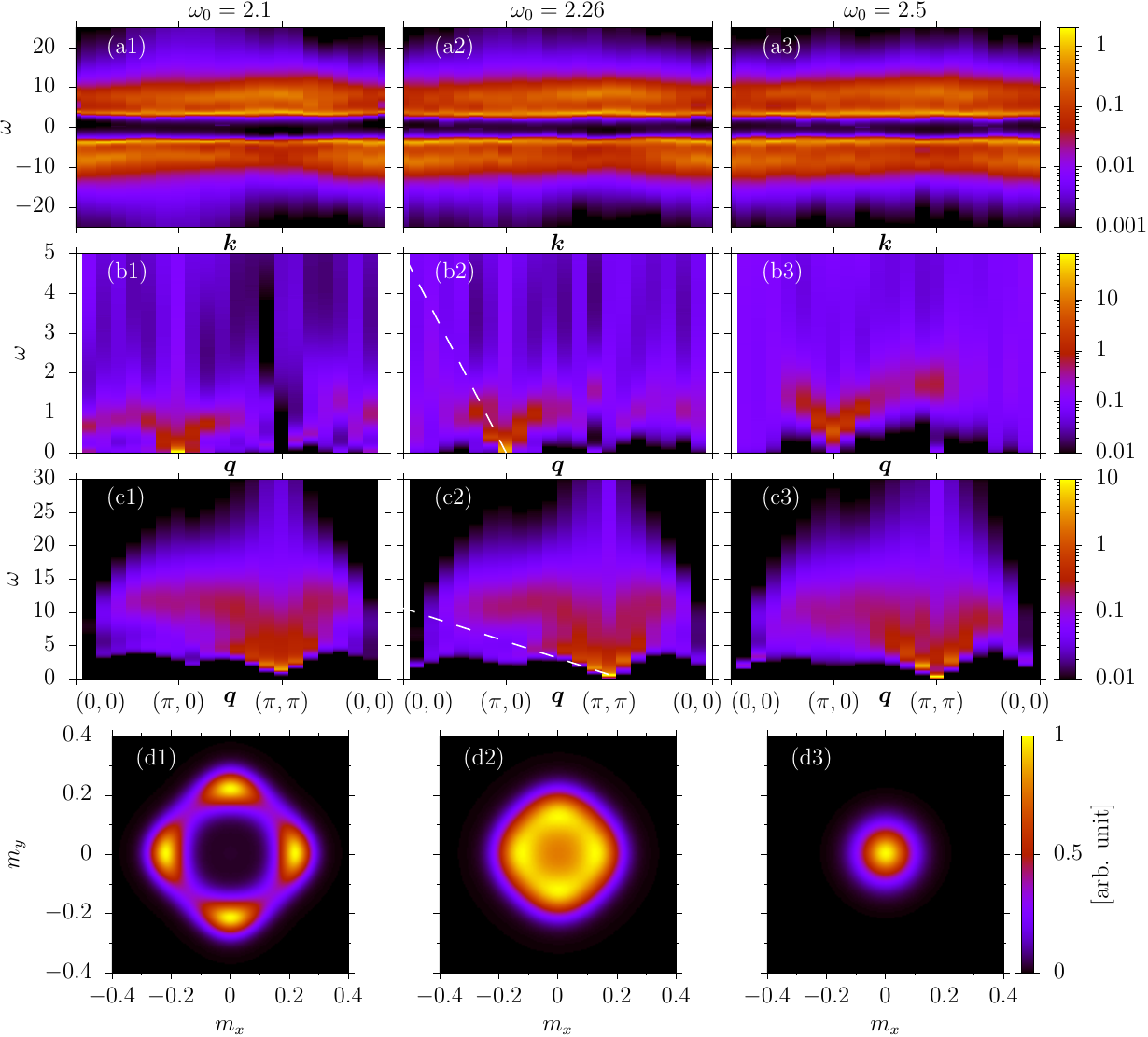}
   \caption{\label{fig:lam075_U05_Spectral}(a1)-(a3) Single-particle spectral function
    $A(\bm{k}, \omega)$, (b1)-(b3) dynamical VBS structure factor
    $S_D(\bm{q},\omega)$, (c1)-(c3) dynamical spin structure factor
    $S_S(\bm{q},\omega)$, and (d1)-(d3) histogram of the VBS order parameter 
    $m_x$ and $m_y$  for different $\omega_0$ and $t=0.1$, $\lambda=0.75$, $U=0.5$,
    $\beta=L=14$.}
\end{figure}

Figure~\ref{fig:lam075_U05_Spectral}  plots  the single-particle, dimer, and spin spectral functions. These quantities are obtained by using the ALF-implementation \cite{ALF_v2} of the stochastic analytical continuation \cite{Beach04a,Sandvik98,Shao23} method. At $t=0$, our  model reduces to an unconstrained $\mathbb{Z}_2$ gauge  theory \cite{Goetz23,Assaad16}.  In this limit, the  electron carries a $\mathbb{Z}_2$  charge, such that the Green function becomes purely local \cite{Goetz23,Nandkishore12,Hohenadler18, Hohenadler19}.  The spectral function at $t=0.1$ in Fig.~\ref{fig:lam075_U05_Spectral}(a) is close to this limit, shows very little dispersion, and a  single-particle gap. The  dimer and spin  dynamical  correlations are shown in Figs.~\ref{fig:lam075_U05_Spectral}(b) and \ref{fig:lam075_U05_Spectral}(c). In  the VBS (AFM) phase the spin (VBS) is gapped.  In the vicinity of the critical point, we  expect emergent Lorentz symmetry that requires  the velocities of the AFM and VBS fluctuations  to be  identical. Within our precision, the  data supports this point of view. Finally, Fig.~\ref{fig:lam075_U05_Spectral}(d) plots the  histogram of the VBS order parameters. In Fig.~\ref{fig:lam075_U05_Spectral}(d2),  we observe a clear sign of emergent  U(1)  symmetry.

In Fig.~\ref{fig:SSHH1}(a), we  show  the evolution of the  spin correlation ratio as a function of  $\lambda$ for $L=\beta = 10$. It can be observed that 
a discontinuity emerges upon  increasing  $\lambda$.  In the supplemental material, we show  that  as for the O(4) model, both $\frac{1}{N_b} \frac{\partial F}{ \partial \omega_0} $  and the histograms of the VBS order  show  the emergence of a first-order transition as $\lambda$ is increased.

\textit{Discussion and conclusions.}
Our  numerical results provide a tuning parameter, the value of the  critical phonon frequency,  that renders the DQCP strongly first order.  Integrating out phonons leads to   retarded interactions in imaginary time,  with  length  scale  set  by  the inverse phonon 
frequency.   Since  the  DQCP enjoys  Lorentz invariance,  enhancing  the range in imaginary times is identical  to  enhancing  it in  real space.  From  this perspective, 
our  results are  consistent  with  the ones observed  in Ref.~\cite{Takahashi24} 
that  show  that  enhancing  the  real-space  range of  the interaction  in 
$J$-$Q_n$  models   results  in 
a  \textit{strong} first-order transition.  Importantly, this tuning parameter does not  alter the symmetries of the model.

The   DQCP is characterized by an emergent compact U(1)  gauge theory  \cite{Senthil04_1,Senthil04_2,Xu18,WangC17}.   On the  square  lattice,  the U(1) symmetry  is  reduced  to $C_4$  symmetry thereby 
allowing  for   quadruple monopole  instances, which  are  understood  to  be  dangerously 
irrelevant.  In Refs.~\cite{Seifert23,Hofmeier24}, it is  argued  that  the  very  same  theory  has 
a Peierls  instability owing to the coupling  of phonons to  monopoles of the emergent U(1)  gauge 
field.  In this  framework,  our results can be understood in terms of a Peierls  instability
of   the compact  U(1) gauge  theory. The critical frequency  at  which this  transition occurs    corresponds to the bullet in Fig.~\ref{fig:sketch}. It is interesting to note  that one can formulate 
monopole  free  realizations of DQC \cite{Liu18}.  Such a model should not  exhibit a Peierls instability and hence this scenario can be tested.   Furthermore,  $J-Q_n$  models where lattice 
fluctuations are absent  show the same phenomena. Hence, the Peierls instability scenario 
cannot be the only interpretation of our results.

There is an  emerging  consensus  that the DQCP is  a   weakly  first-order  transition.
The assumption  that the DQCP is a  critical point places  strong  constraints on the 
correlation length exponent  \cite{Nakayama16} that seem at  odds  with Monte Carlo  
results \cite{Shao15,Nahum15}.    Furthermore, 
there has  been considerable numerical evidence  for  emergent SO(5) symmetry \cite{Nahum15_1,SatoT22}. 
One possible  way of  understanding this  is in  terms  of  fixed point annihilation 
\cite{WangC17}  resulting  in a  non-unitary SO(5)-CFT \cite{WangZ20}.  
Assuming  that the 
complex   fixed point is  close  to the  real plane,  then  the  flow  in the  proximity 
of  this   fixed point  will  be  very  slow.
In this  scenario, models  with short range interactions  would exhibit an RG flow that approaches the 
complex critical point. As such, very large system sizes  are required to detect the first order nature of the transition.   Models  with  longer ranged interactions  would  exhibit flows  that flow away from the fixed point.  Thereby, much smaller system sizes are required  to resolve  the  first order nature  of the 
transition.   Following  this point of view,  the bullet point in Fig.~\ref{fig:sketch} corresponds  to a complex fixed point in the proximity of which the RG flow is  very slow, but  always flows in one direction. 

The aforementioned bound on the  correlation length exponent holds only for a  critical point that, by definition, has a single relevant operator. It does not prohibit   an understanding of  the DQCP in
terms of  an SO(5)  multi-critical  point,  where  the leading  SO(5)  singlet  operator is  relevant \cite{Chester24}. This is the  point of  view  put  forward  in   \cite{Takahashi24}.  In this  context,  enhancing  $\lambda$   tunes  away  from the  multi-critical point  resulting in 
a  strong  first-order  transition.  In this reading of  our data,  the bullet point in Fig.~\ref{fig:sketch} corresponds  to  an   SO(5) multi-critical point. 

For a generic critical point with no dangerously irrelevant operators, a small symmetry preserving change in the range 
of the interaction---that merely corresponds to a different lattice regularization of the relevant operator---should not change the criticality. The fact that we see a big change in criticality even under a small change of the critical  frequency points to the very special nature of DQC.  Our results hold for both the O(4) as well as the generic $\mathrm{SU}(2)\times C_4$  realizations of DQC. The interaction range provides the missing tuning parameter that must exist in the  complex CFT or in the multi-critical interpretations of DQCP.

\textit{Acknowledgments.}
  We would like to thank discussions  with J. Willsher  and  J. Carvalho In\'acio. The authors gratefully acknowledge the Gauss Centre for Supercomputing
  e.V. (www.gauss-centre.eu) for funding this project by providing computing
  time on the GCS Supercomputer SuperMUC-NG at the Leibniz Supercomputing Centre
  (www.lrz.de).  
  The authors gratefully acknowledge the scientific support and HPC resources provided by the Erlangen National High Performance Computing Center (NHR@FAU) of the Friedrich-Alexander-Universit\"at Erlangen-N\"urnberg (FAU) under NHR Project No. 80069. NHR funding is provided by federal and Bavarian state authorities. NHR@FAU hardware is partially funded by the German Research Foundation (DFG) through Grant No. 440719683.
  F.F.A. thanks the W\"urzburg-Dresden
  Cluster of Excellence on Complexity and Topology in Quantum Matter ct.qmat
  (EXC 2147, project-id 390858490),  and A.G. acknowledges the DFG funded SFB 1170 on Topological
  and Correlated Electronics at Surfaces and Interfaces (Project No. 258499086).
N.C.C~is grateful to the Brazilian Agencies Conselho Nacional de Desenvolvimento Cientif\'ico e
Tecnol\'ogico (CNPq), Coordena\c{c}\~ao de Aperfei\c coamento de Pessoal de Ensino Superior (CAPES), and Funda\c{c}\~ao de Amparo \`a Pesquisa do Estado do Rio de Janeiro, FAPERJ.
N.C.C.~acknowledges support from FAPERJ Grant No.~E-26/200.258/2023 - SEI-260003/000623/2023, and CNPq Grant No.~313065/2021-7.
\bibliography{fassaad}

\begin{thebibliography}{41}%
\makeatletter
\providecommand \@ifxundefined [1]{%
 \@ifx{#1\undefined}
}%
\providecommand \@ifnum [1]{%
 \ifnum #1\expandafter \@firstoftwo
 \else \expandafter \@secondoftwo
 \fi
}%
\providecommand \@ifx [1]{%
 \ifx #1\expandafter \@firstoftwo
 \else \expandafter \@secondoftwo
 \fi
}%
\providecommand \natexlab [1]{#1}%
\providecommand \enquote  [1]{``#1''}%
\providecommand \bibnamefont  [1]{#1}%
\providecommand \bibfnamefont [1]{#1}%
\providecommand \citenamefont [1]{#1}%
\providecommand \href@noop [0]{\@secondoftwo}%
\providecommand \href [0]{\begingroup \@sanitize@url \@href}%
\providecommand \@href[1]{\@@startlink{#1}\@@href}%
\providecommand \@@href[1]{\endgroup#1\@@endlink}%
\providecommand \@sanitize@url [0]{\catcode `\\12\catcode `\$12\catcode
  `\&12\catcode `\#12\catcode `\^12\catcode `\_12\catcode `\%12\relax}%
\providecommand \@@startlink[1]{}%
\providecommand \@@endlink[0]{}%
\providecommand \url  [0]{\begingroup\@sanitize@url \@url }%
\providecommand \@url [1]{\endgroup\@href {#1}{\urlprefix }}%
\providecommand \urlprefix  [0]{URL }%
\providecommand \Eprint [0]{\href }%
\providecommand \doibase [0]{https://doi.org/}%
\providecommand \selectlanguage [0]{\@gobble}%
\providecommand \bibinfo  [0]{\@secondoftwo}%
\providecommand \bibfield  [0]{\@secondoftwo}%
\providecommand \translation [1]{[#1]}%
\providecommand \BibitemOpen [0]{}%
\providecommand \bibitemStop [0]{}%
\providecommand \bibitemNoStop [0]{.\EOS\space}%
\providecommand \EOS [0]{\spacefactor3000\relax}%
\providecommand \BibitemShut  [1]{\csname bibitem#1\endcsname}%
\let\auto@bib@innerbib\@empty
%</preamble>
\bibitem [{\citenamefont {Senthil}\ \emph
  {et~al.}(2004{\natexlab{a}})\citenamefont {Senthil}, \citenamefont {Balents},
  \citenamefont {Sachdev}, \citenamefont {Vishwanath},\ and\ \citenamefont
  {Fisher}}]{Senthil04_1}%
  \BibitemOpen
  \bibfield  {author} {\bibinfo {author} {\bibfnamefont {T.}~\bibnamefont
  {Senthil}}, \bibinfo {author} {\bibfnamefont {L.}~\bibnamefont {Balents}},
  \bibinfo {author} {\bibfnamefont {S.}~\bibnamefont {Sachdev}}, \bibinfo
  {author} {\bibfnamefont {A.}~\bibnamefont {Vishwanath}},\ and\ \bibinfo
  {author} {\bibfnamefont {M.~P.~A.}\ \bibnamefont {Fisher}},\ }\bibfield
  {title} {\bibinfo {title} {Quantum criticality beyond the
  landau-ginzburg-wilson paradigm},\ }\href
  {https://doi.org/10.1103/PhysRevB.70.144407} {\bibfield  {journal} {\bibinfo
  {journal} {Phys. Rev. B}\ }\textbf {\bibinfo {volume} {70}},\ \bibinfo
  {pages} {144407} (\bibinfo {year} {2004}{\natexlab{a}})}\BibitemShut
  {NoStop}%
\bibitem [{\citenamefont {Senthil}\ \emph
  {et~al.}(2004{\natexlab{b}})\citenamefont {Senthil}, \citenamefont
  {Vishwanath}, \citenamefont {Balents}, \citenamefont {Sachdev},\ and\
  \citenamefont {Fisher}}]{Senthil04_2}%
  \BibitemOpen
  \bibfield  {author} {\bibinfo {author} {\bibfnamefont {T.}~\bibnamefont
  {Senthil}}, \bibinfo {author} {\bibfnamefont {A.}~\bibnamefont {Vishwanath}},
  \bibinfo {author} {\bibfnamefont {L.}~\bibnamefont {Balents}}, \bibinfo
  {author} {\bibfnamefont {S.}~\bibnamefont {Sachdev}},\ and\ \bibinfo {author}
  {\bibfnamefont {M.~P.~A.}\ \bibnamefont {Fisher}},\ }\bibfield  {title}
  {\bibinfo {title} {Deconfined quantum critical points},\ }\href
  {https://doi.org/10.1126/science.1091806} {\bibfield  {journal} {\bibinfo
  {journal} {Science}\ }\textbf {\bibinfo {volume} {303}},\ \bibinfo {pages}
  {1490} (\bibinfo {year} {2004}{\natexlab{b}})}\BibitemShut {NoStop}%
\bibitem [{\citenamefont {Wang}\ \emph {et~al.}(2017)\citenamefont {Wang},
  \citenamefont {Nahum}, \citenamefont {Metlitski}, \citenamefont {Xu},\ and\
  \citenamefont {Senthil}}]{WangC17}%
  \BibitemOpen
  \bibfield  {author} {\bibinfo {author} {\bibfnamefont {C.}~\bibnamefont
  {Wang}}, \bibinfo {author} {\bibfnamefont {A.}~\bibnamefont {Nahum}},
  \bibinfo {author} {\bibfnamefont {M.~A.}\ \bibnamefont {Metlitski}}, \bibinfo
  {author} {\bibfnamefont {C.}~\bibnamefont {Xu}},\ and\ \bibinfo {author}
  {\bibfnamefont {T.}~\bibnamefont {Senthil}},\ }\bibfield  {title} {\bibinfo
  {title} {Deconfined quantum critical points: Symmetries and dualities},\
  }\href {https://doi.org/10.1103/PhysRevX.7.031051} {\bibfield  {journal}
  {\bibinfo  {journal} {Phys. Rev. X}\ }\textbf {\bibinfo {volume} {7}},\
  \bibinfo {pages} {031051} (\bibinfo {year} {2017})}\BibitemShut {NoStop}%
\bibitem [{\citenamefont {Tanaka}\ and\ \citenamefont {Hu}(2005)}]{Tanaka05}%
  \BibitemOpen
  \bibfield  {author} {\bibinfo {author} {\bibfnamefont {A.}~\bibnamefont
  {Tanaka}}\ and\ \bibinfo {author} {\bibfnamefont {X.}~\bibnamefont {Hu}},\
  }\bibfield  {title} {\bibinfo {title} {Many-body spin berry phases emerging
  from the $\ensuremath{\pi}$-flux state: Competition between
  antiferromagnetism and the valence-bond-solid state},\ }\href
  {https://doi.org/10.1103/PhysRevLett.95.036402} {\bibfield  {journal}
  {\bibinfo  {journal} {Phys. Rev. Lett.}\ }\textbf {\bibinfo {volume} {95}},\
  \bibinfo {pages} {036402} (\bibinfo {year} {2005})}\BibitemShut {NoStop}%
\bibitem [{\citenamefont {Abanov}\ and\ \citenamefont
  {Wiegmann}(2000)}]{Abanov00}%
  \BibitemOpen
  \bibfield  {author} {\bibinfo {author} {\bibfnamefont {A.}~\bibnamefont
  {Abanov}}\ and\ \bibinfo {author} {\bibfnamefont {P.}~\bibnamefont
  {Wiegmann}},\ }\bibfield  {title} {\bibinfo {title} {Theta-terms in nonlinear
  sigma-models},\ }\href
  {https://doi.org/http://dx.doi.org/10.1016/S0550-3213(99)00820-2} {\bibfield
  {journal} {\bibinfo  {journal} {Nuclear Physics B}\ }\textbf {\bibinfo
  {volume} {570}},\ \bibinfo {pages} {685 } (\bibinfo {year}
  {2000})}\BibitemShut {NoStop}%
\bibitem [{\citenamefont {Senthil}\ and\ \citenamefont
  {Fisher}(2006)}]{Senthil06}%
  \BibitemOpen
  \bibfield  {author} {\bibinfo {author} {\bibfnamefont {T.}~\bibnamefont
  {Senthil}}\ and\ \bibinfo {author} {\bibfnamefont {M.~P.~A.}\ \bibnamefont
  {Fisher}},\ }\bibfield  {title} {\bibinfo {title} {Competing orders,
  nonlinear sigma models, and topological terms in quantum magnets},\ }\href
  {https://doi.org/10.1103/PhysRevB.74.064405} {\bibfield  {journal} {\bibinfo
  {journal} {Phys. Rev. B}\ }\textbf {\bibinfo {volume} {74}},\ \bibinfo
  {pages} {064405} (\bibinfo {year} {2006})}\BibitemShut {NoStop}%
\bibitem [{\citenamefont {Ryu}\ \emph {et~al.}(2009)\citenamefont {Ryu},
  \citenamefont {Mudry}, \citenamefont {Hou},\ and\ \citenamefont
  {Chamon}}]{Ryu09}%
  \BibitemOpen
  \bibfield  {author} {\bibinfo {author} {\bibfnamefont {S.}~\bibnamefont
  {Ryu}}, \bibinfo {author} {\bibfnamefont {C.}~\bibnamefont {Mudry}}, \bibinfo
  {author} {\bibfnamefont {C.-Y.}\ \bibnamefont {Hou}},\ and\ \bibinfo {author}
  {\bibfnamefont {C.}~\bibnamefont {Chamon}},\ }\bibfield  {title} {\bibinfo
  {title} {Masses in graphenelike two-dimensional electronic systems:
  Topological defects in order parameters and their fractional exchange
  statistics},\ }\href {https://doi.org/10.1103/PhysRevB.80.205319} {\bibfield
  {journal} {\bibinfo  {journal} {Phys. Rev. B}\ }\textbf {\bibinfo {volume}
  {80}},\ \bibinfo {pages} {205319} (\bibinfo {year} {2009})}\BibitemShut
  {NoStop}%
\bibitem [{\citenamefont {Levin}\ and\ \citenamefont
  {Senthil}(2004)}]{Levin04}%
  \BibitemOpen
  \bibfield  {author} {\bibinfo {author} {\bibfnamefont {M.}~\bibnamefont
  {Levin}}\ and\ \bibinfo {author} {\bibfnamefont {T.}~\bibnamefont
  {Senthil}},\ }\bibfield  {title} {\bibinfo {title} {Deconfined quantum
  criticality and n\'eel order via dimer disorder},\ }\href
  {https://doi.org/10.1103/PhysRevB.70.220403} {\bibfield  {journal} {\bibinfo
  {journal} {Phys. Rev. B}\ }\textbf {\bibinfo {volume} {70}},\ \bibinfo
  {pages} {220403} (\bibinfo {year} {2004})}\BibitemShut {NoStop}%
\bibitem [{\citenamefont {Liu}\ \emph {et~al.}(2022)\citenamefont {Liu},
  \citenamefont {Vojta}, \citenamefont {Assaad},\ and\ \citenamefont
  {Janssen}}]{LiuZH21}%
  \BibitemOpen
  \bibfield  {author} {\bibinfo {author} {\bibfnamefont {Z.~H.}\ \bibnamefont
  {Liu}}, \bibinfo {author} {\bibfnamefont {M.}~\bibnamefont {Vojta}}, \bibinfo
  {author} {\bibfnamefont {F.~F.}\ \bibnamefont {Assaad}},\ and\ \bibinfo
  {author} {\bibfnamefont {L.}~\bibnamefont {Janssen}},\ }\bibfield  {title}
  {\bibinfo {title} {Metallic and deconfined quantum criticality in dirac
  systems},\ }\href {https://doi.org/10.1103/PhysRevLett.128.087201} {\bibfield
   {journal} {\bibinfo  {journal} {Phys. Rev. Lett.}\ }\textbf {\bibinfo
  {volume} {128}},\ \bibinfo {pages} {087201} (\bibinfo {year}
  {2022})}\BibitemShut {NoStop}%
\bibitem [{\citenamefont {Liu}\ \emph {et~al.}(2024)\citenamefont {Liu},
  \citenamefont {Vojta}, \citenamefont {Assaad},\ and\ \citenamefont
  {Janssen}}]{LiuZ24}%
  \BibitemOpen
  \bibfield  {author} {\bibinfo {author} {\bibfnamefont {Z.~H.}\ \bibnamefont
  {Liu}}, \bibinfo {author} {\bibfnamefont {M.}~\bibnamefont {Vojta}}, \bibinfo
  {author} {\bibfnamefont {F.~F.}\ \bibnamefont {Assaad}},\ and\ \bibinfo
  {author} {\bibfnamefont {L.}~\bibnamefont {Janssen}},\ }\bibfield  {title}
  {\bibinfo {title} {Critical properties of metallic and deconfined quantum
  phase transitions in dirac systems},\ }\href
  {https://doi.org/10.1103/PhysRevB.110.125123} {\bibfield  {journal} {\bibinfo
   {journal} {Phys. Rev. B}\ }\textbf {\bibinfo {volume} {110}},\ \bibinfo
  {pages} {125123} (\bibinfo {year} {2024})}\BibitemShut {NoStop}%
\bibitem [{\citenamefont {Sandvik}(2007)}]{Sandvik07}%
  \BibitemOpen
  \bibfield  {author} {\bibinfo {author} {\bibfnamefont {A.~W.}\ \bibnamefont
  {Sandvik}},\ }\bibfield  {title} {\bibinfo {title} {Evidence for deconfined
  quantum criticality in a two-dimensional heisenberg model with four-spin
  interactions},\ }\href {https://doi.org/10.1103/PhysRevLett.98.227202}
  {\bibfield  {journal} {\bibinfo  {journal} {Phys. Rev. Lett.}\ }\textbf
  {\bibinfo {volume} {98}},\ \bibinfo {pages} {227202} (\bibinfo {year}
  {2007})}\BibitemShut {NoStop}%
\bibitem [{\citenamefont {Nahum}\ \emph
  {et~al.}(2015{\natexlab{a}})\citenamefont {Nahum}, \citenamefont {Chalker},
  \citenamefont {Serna}, \citenamefont {Ortu\~no},\ and\ \citenamefont
  {Somoza}}]{Nahum15}%
  \BibitemOpen
  \bibfield  {author} {\bibinfo {author} {\bibfnamefont {A.}~\bibnamefont
  {Nahum}}, \bibinfo {author} {\bibfnamefont {J.~T.}\ \bibnamefont {Chalker}},
  \bibinfo {author} {\bibfnamefont {P.}~\bibnamefont {Serna}}, \bibinfo
  {author} {\bibfnamefont {M.}~\bibnamefont {Ortu\~no}},\ and\ \bibinfo
  {author} {\bibfnamefont {A.~M.}\ \bibnamefont {Somoza}},\ }\bibfield  {title}
  {\bibinfo {title} {Deconfined quantum criticality, scaling violations, and
  classical loop models},\ }\href {https://doi.org/10.1103/PhysRevX.5.041048}
  {\bibfield  {journal} {\bibinfo  {journal} {Phys. Rev. X}\ }\textbf {\bibinfo
  {volume} {5}},\ \bibinfo {pages} {041048} (\bibinfo {year}
  {2015}{\natexlab{a}})}\BibitemShut {NoStop}%
\bibitem [{\citenamefont {Liu}\ \emph {et~al.}(2019)\citenamefont {Liu},
  \citenamefont {Wang}, \citenamefont {Sato}, \citenamefont {Hohenadler},
  \citenamefont {Wang}, \citenamefont {Guo},\ and\ \citenamefont
  {Assaad}}]{Liu18}%
  \BibitemOpen
  \bibfield  {author} {\bibinfo {author} {\bibfnamefont {Y.}~\bibnamefont
  {Liu}}, \bibinfo {author} {\bibfnamefont {Z.}~\bibnamefont {Wang}}, \bibinfo
  {author} {\bibfnamefont {T.}~\bibnamefont {Sato}}, \bibinfo {author}
  {\bibfnamefont {M.}~\bibnamefont {Hohenadler}}, \bibinfo {author}
  {\bibfnamefont {C.}~\bibnamefont {Wang}}, \bibinfo {author} {\bibfnamefont
  {W.}~\bibnamefont {Guo}},\ and\ \bibinfo {author} {\bibfnamefont {F.~F.}\
  \bibnamefont {Assaad}},\ }\bibfield  {title} {\bibinfo {title}
  {{Superconductivity from the condensation of topological defects in a quantum
  spin-Hall insulator}},\ }\href {https://doi.org/10.1038/s41467-019-10372-0}
  {\bibfield  {journal} {\bibinfo  {journal} {Nature Communications}\ }\textbf
  {\bibinfo {volume} {10}},\ \bibinfo {pages} {2658} (\bibinfo {year}
  {2019})}\BibitemShut {NoStop}%
\bibitem [{\citenamefont {G\"otz}\ \emph {et~al.}(2024)\citenamefont {G\"otz},
  \citenamefont {Hohenadler},\ and\ \citenamefont {Assaad}}]{Goetz23}%
  \BibitemOpen
  \bibfield  {author} {\bibinfo {author} {\bibfnamefont {A.}~\bibnamefont
  {G\"otz}}, \bibinfo {author} {\bibfnamefont {M.}~\bibnamefont {Hohenadler}},\
  and\ \bibinfo {author} {\bibfnamefont {F.~F.}\ \bibnamefont {Assaad}},\
  }\bibfield  {title} {\bibinfo {title} {Phases and exotic phase transitions of
  a two-dimensional su-schrieffer-heeger model},\ }\href
  {https://doi.org/10.1103/PhysRevB.109.195154} {\bibfield  {journal} {\bibinfo
   {journal} {Phys. Rev. B}\ }\textbf {\bibinfo {volume} {109}},\ \bibinfo
  {pages} {195154} (\bibinfo {year} {2024})}\BibitemShut {NoStop}%
\bibitem [{\citenamefont {Nahum}\ \emph
  {et~al.}(2015{\natexlab{b}})\citenamefont {Nahum}, \citenamefont {Serna},
  \citenamefont {Chalker}, \citenamefont {Ortu\~no},\ and\ \citenamefont
  {Somoza}}]{Nahum15_1}%
  \BibitemOpen
  \bibfield  {author} {\bibinfo {author} {\bibfnamefont {A.}~\bibnamefont
  {Nahum}}, \bibinfo {author} {\bibfnamefont {P.}~\bibnamefont {Serna}},
  \bibinfo {author} {\bibfnamefont {J.~T.}\ \bibnamefont {Chalker}}, \bibinfo
  {author} {\bibfnamefont {M.}~\bibnamefont {Ortu\~no}},\ and\ \bibinfo
  {author} {\bibfnamefont {A.~M.}\ \bibnamefont {Somoza}},\ }\bibfield  {title}
  {\bibinfo {title} {Emergent so(5) symmetry at the n\'eel to
  valence-bond-solid transition},\ }\href
  {https://doi.org/10.1103/PhysRevLett.115.267203} {\bibfield  {journal}
  {\bibinfo  {journal} {Phys. Rev. Lett.}\ }\textbf {\bibinfo {volume} {115}},\
  \bibinfo {pages} {267203} (\bibinfo {year} {2015}{\natexlab{b}})}\BibitemShut
  {NoStop}%
\bibitem [{\citenamefont {Sato}\ \emph {et~al.}(2023)\citenamefont {Sato},
  \citenamefont {Wang}, \citenamefont {Liu}, \citenamefont {Hou}, \citenamefont
  {Hohenadler}, \citenamefont {Guo},\ and\ \citenamefont {Assaad}}]{SatoT22}%
  \BibitemOpen
  \bibfield  {author} {\bibinfo {author} {\bibfnamefont {T.}~\bibnamefont
  {Sato}}, \bibinfo {author} {\bibfnamefont {Z.}~\bibnamefont {Wang}}, \bibinfo
  {author} {\bibfnamefont {Y.}~\bibnamefont {Liu}}, \bibinfo {author}
  {\bibfnamefont {D.}~\bibnamefont {Hou}}, \bibinfo {author} {\bibfnamefont
  {M.}~\bibnamefont {Hohenadler}}, \bibinfo {author} {\bibfnamefont
  {W.}~\bibnamefont {Guo}},\ and\ \bibinfo {author} {\bibfnamefont {F.~F.}\
  \bibnamefont {Assaad}},\ }\bibfield  {title} {\bibinfo {title} {Simulation of
  fermionic and bosonic critical points with emergent so(5) symmetry},\ }\href
  {https://doi.org/10.1103/PhysRevB.108.L121111} {\bibfield  {journal}
  {\bibinfo  {journal} {Phys. Rev. B}\ }\textbf {\bibinfo {volume} {108}},\
  \bibinfo {pages} {L121111} (\bibinfo {year} {2023})}\BibitemShut {NoStop}%
\bibitem [{\citenamefont {Nakayama}\ and\ \citenamefont
  {Ohtsuki}(2016)}]{Nakayama16}%
  \BibitemOpen
  \bibfield  {author} {\bibinfo {author} {\bibfnamefont {Y.}~\bibnamefont
  {Nakayama}}\ and\ \bibinfo {author} {\bibfnamefont {T.}~\bibnamefont
  {Ohtsuki}},\ }\bibfield  {title} {\bibinfo {title} {Necessary condition for
  emergent symmetry from the conformal bootstrap},\ }\href
  {https://doi.org/10.1103/PhysRevLett.117.131601} {\bibfield  {journal}
  {\bibinfo  {journal} {Phys. Rev. Lett.}\ }\textbf {\bibinfo {volume} {117}},\
  \bibinfo {pages} {131601} (\bibinfo {year} {2016})}\BibitemShut {NoStop}%
\bibitem [{\citenamefont {Ma}\ and\ \citenamefont {Wang}(2020)}]{WangC19}%
  \BibitemOpen
  \bibfield  {author} {\bibinfo {author} {\bibfnamefont {R.}~\bibnamefont
  {Ma}}\ and\ \bibinfo {author} {\bibfnamefont {C.}~\bibnamefont {Wang}},\
  }\bibfield  {title} {\bibinfo {title} {Theory of deconfined
  pseudocriticality},\ }\href {https://doi.org/10.1103/PhysRevB.102.020407}
  {\bibfield  {journal} {\bibinfo  {journal} {Phys. Rev. B}\ }\textbf {\bibinfo
  {volume} {102}},\ \bibinfo {pages} {020407} (\bibinfo {year}
  {2020})}\BibitemShut {NoStop}%
\bibitem [{\citenamefont {Nahum}(2020)}]{Nahum19}%
  \BibitemOpen
  \bibfield  {author} {\bibinfo {author} {\bibfnamefont {A.}~\bibnamefont
  {Nahum}},\ }\bibfield  {title} {\bibinfo {title} {Note on wess-zumino-witten
  models and quasiuniversality in $2+1$ dimensions},\ }\href
  {https://doi.org/10.1103/PhysRevB.102.201116} {\bibfield  {journal} {\bibinfo
   {journal} {Phys. Rev. B}\ }\textbf {\bibinfo {volume} {102}},\ \bibinfo
  {pages} {201116} (\bibinfo {year} {2020})}\BibitemShut {NoStop}%
\bibitem [{\citenamefont {Chester}\ and\ \citenamefont {Su}(2024)}]{Chester24}%
  \BibitemOpen
  \bibfield  {author} {\bibinfo {author} {\bibfnamefont {S.~M.}\ \bibnamefont
  {Chester}}\ and\ \bibinfo {author} {\bibfnamefont {N.}~\bibnamefont {Su}},\
  }\bibfield  {title} {\bibinfo {title} {Bootstrapping deconfined quantum
  tricriticality},\ }\href {https://doi.org/10.1103/PhysRevLett.132.111601}
  {\bibfield  {journal} {\bibinfo  {journal} {Phys. Rev. Lett.}\ }\textbf
  {\bibinfo {volume} {132}},\ \bibinfo {pages} {111601} (\bibinfo {year}
  {2024})}\BibitemShut {NoStop}%
\bibitem [{\citenamefont {Takahashi}\ \emph {et~al.}(2024)\citenamefont
  {Takahashi}, \citenamefont {Shao}, \citenamefont {Zhao}, \citenamefont
  {Guo},\ and\ \citenamefont {Sandvik}}]{Takahashi24}%
  \BibitemOpen
  \bibfield  {author} {\bibinfo {author} {\bibfnamefont {J.}~\bibnamefont
  {Takahashi}}, \bibinfo {author} {\bibfnamefont {H.}~\bibnamefont {Shao}},
  \bibinfo {author} {\bibfnamefont {B.}~\bibnamefont {Zhao}}, \bibinfo {author}
  {\bibfnamefont {W.}~\bibnamefont {Guo}},\ and\ \bibinfo {author}
  {\bibfnamefont {A.~W.}\ \bibnamefont {Sandvik}},\ }\bibfield  {title}
  {\bibinfo {title} {So(5) multicriticality in two-dimensional quantum
  magnets},\ }\href@noop {} {\bibfield  {journal} {\bibinfo  {journal}
  {arXiv:2405.06607}\ } (\bibinfo {year} {2024})},\ \Eprint
  {https://arxiv.org/abs/2405.06607} {arXiv:2405.06607 [cond-mat.str-el]}
  \BibitemShut {NoStop}%
\bibitem [{\citenamefont {Su}\ \emph {et~al.}(1980)\citenamefont {Su},
  \citenamefont {Schrieffer},\ and\ \citenamefont {Heeger}}]{Su80}%
  \BibitemOpen
  \bibfield  {author} {\bibinfo {author} {\bibfnamefont {W.~P.}\ \bibnamefont
  {Su}}, \bibinfo {author} {\bibfnamefont {J.~R.}\ \bibnamefont {Schrieffer}},\
  and\ \bibinfo {author} {\bibfnamefont {A.~J.}\ \bibnamefont {Heeger}},\
  }\bibfield  {title} {\bibinfo {title} {Soliton excitations in
  polyacetylene},\ }\href {https://doi.org/10.1103/PhysRevB.22.2099} {\bibfield
   {journal} {\bibinfo  {journal} {Phys. Rev. B}\ }\textbf {\bibinfo {volume}
  {22}},\ \bibinfo {pages} {2099} (\bibinfo {year} {1980})}\BibitemShut
  {NoStop}%
\bibitem [{\citenamefont {Seifert}\ \emph {et~al.}(2024)\citenamefont
  {Seifert}, \citenamefont {Willsher}, \citenamefont {Drescher}, \citenamefont
  {Pollmann},\ and\ \citenamefont {Knolle}}]{Seifert23}%
  \BibitemOpen
  \bibfield  {author} {\bibinfo {author} {\bibfnamefont {U.~F.~P.}\
  \bibnamefont {Seifert}}, \bibinfo {author} {\bibfnamefont {J.}~\bibnamefont
  {Willsher}}, \bibinfo {author} {\bibfnamefont {M.}~\bibnamefont {Drescher}},
  \bibinfo {author} {\bibfnamefont {F.}~\bibnamefont {Pollmann}},\ and\
  \bibinfo {author} {\bibfnamefont {J.}~\bibnamefont {Knolle}},\ }\bibfield
  {title} {\bibinfo {title} {Spin-peierls instability of the u(1) dirac spin
  liquid},\ }\href {https://doi.org/10.1038/s41467-024-51367-w} {\bibfield
  {journal} {\bibinfo  {journal} {Nature Communications}\ }\textbf {\bibinfo
  {volume} {15}},\ \bibinfo {pages} {7110} (\bibinfo {year}
  {2024})}\BibitemShut {NoStop}%
\bibitem [{\citenamefont {Hofmeier}\ \emph {et~al.}(2024)\citenamefont
  {Hofmeier}, \citenamefont {Willsher}, \citenamefont {Seifert},\ and\
  \citenamefont {Knolle}}]{Hofmeier24}%
  \BibitemOpen
  \bibfield  {author} {\bibinfo {author} {\bibfnamefont {D.}~\bibnamefont
  {Hofmeier}}, \bibinfo {author} {\bibfnamefont {J.}~\bibnamefont {Willsher}},
  \bibinfo {author} {\bibfnamefont {U.~F.~P.}\ \bibnamefont {Seifert}},\ and\
  \bibinfo {author} {\bibfnamefont {J.}~\bibnamefont {Knolle}},\ }\bibfield
  {title} {\bibinfo {title} {Spin-peierls instability of deconfined quantum
  critical points},\ }\href {https://doi.org/10.1103/PhysRevB.110.125130}
  {\bibfield  {journal} {\bibinfo  {journal} {Phys. Rev. B}\ }\textbf {\bibinfo
  {volume} {110}},\ \bibinfo {pages} {125130} (\bibinfo {year}
  {2024})}\BibitemShut {NoStop}%
\bibitem [{\citenamefont {Wang}\ \emph {et~al.}(2021)\citenamefont {Wang},
  \citenamefont {Zaletel}, \citenamefont {Mong},\ and\ \citenamefont
  {Assaad}}]{WangZ20}%
  \BibitemOpen
  \bibfield  {author} {\bibinfo {author} {\bibfnamefont {Z.}~\bibnamefont
  {Wang}}, \bibinfo {author} {\bibfnamefont {M.~P.}\ \bibnamefont {Zaletel}},
  \bibinfo {author} {\bibfnamefont {R.~S.~K.}\ \bibnamefont {Mong}},\ and\
  \bibinfo {author} {\bibfnamefont {F.~F.}\ \bibnamefont {Assaad}},\ }\bibfield
   {title} {\bibinfo {title} {{Phases of the ($2+1$) Dimensional SO(5)
  Nonlinear Sigma Model with Topological Term}},\ }\href
  {https://doi.org/10.1103/PhysRevLett.126.045701} {\bibfield  {journal}
  {\bibinfo  {journal} {Phys. Rev. Lett.}\ }\textbf {\bibinfo {volume} {126}},\
  \bibinfo {pages} {045701} (\bibinfo {year} {2021})}\BibitemShut {NoStop}%
\bibitem [{\citenamefont {Karakuzu}\ \emph {et~al.}(2018)\citenamefont
  {Karakuzu}, \citenamefont {Seki},\ and\ \citenamefont
  {Sorella}}]{Karakuzu18}%
  \BibitemOpen
  \bibfield  {author} {\bibinfo {author} {\bibfnamefont {S.}~\bibnamefont
  {Karakuzu}}, \bibinfo {author} {\bibfnamefont {K.}~\bibnamefont {Seki}},\
  and\ \bibinfo {author} {\bibfnamefont {S.}~\bibnamefont {Sorella}},\
  }\bibfield  {title} {\bibinfo {title} {Solution of the sign problem for the
  half-filled hubbard-holstein model},\ }\href
  {https://doi.org/10.1103/PhysRevB.98.201108} {\bibfield  {journal} {\bibinfo
  {journal} {Phys. Rev. B}\ }\textbf {\bibinfo {volume} {98}},\ \bibinfo
  {pages} {201108} (\bibinfo {year} {2018})}\BibitemShut {NoStop}%
\bibitem [{\citenamefont {Bercx}\ \emph {et~al.}(2017)\citenamefont {Bercx},
  \citenamefont {Goth}, \citenamefont {Hofmann},\ and\ \citenamefont
  {Assaad}}]{ALF_v1}%
  \BibitemOpen
  \bibfield  {author} {\bibinfo {author} {\bibfnamefont {M.}~\bibnamefont
  {Bercx}}, \bibinfo {author} {\bibfnamefont {F.}~\bibnamefont {Goth}},
  \bibinfo {author} {\bibfnamefont {J.~S.}\ \bibnamefont {Hofmann}},\ and\
  \bibinfo {author} {\bibfnamefont {F.~F.}\ \bibnamefont {Assaad}},\ }\bibfield
   {title} {\bibinfo {title} {{The ALF (Algorithms for Lattice Fermions)
  project release 1.0. Documentation for the auxiliary field quantum Monte
  Carlo code}},\ }\href {https://doi.org/10.21468/SciPostPhys.3.2.013}
  {\bibfield  {journal} {\bibinfo  {journal} {SciPost Phys.}\ }\textbf
  {\bibinfo {volume} {3}},\ \bibinfo {pages} {013} (\bibinfo {year}
  {2017})}\BibitemShut {NoStop}%
\bibitem [{\citenamefont {Assaad}\ \emph {et~al.}(2022)\citenamefont {Assaad},
  \citenamefont {Bercx}, \citenamefont {Goth}, \citenamefont {G{\"o}tz},
  \citenamefont {Hofmann}, \citenamefont {Huffman}, \citenamefont {Liu},
  \citenamefont {Toldin}, \citenamefont {Portela},\ and\ \citenamefont
  {Schwab}}]{ALF_v2}%
  \BibitemOpen
  \bibfield  {author} {\bibinfo {author} {\bibfnamefont {F.~F.}\ \bibnamefont
  {Assaad}}, \bibinfo {author} {\bibfnamefont {M.}~\bibnamefont {Bercx}},
  \bibinfo {author} {\bibfnamefont {F.}~\bibnamefont {Goth}}, \bibinfo {author}
  {\bibfnamefont {A.}~\bibnamefont {G{\"o}tz}}, \bibinfo {author}
  {\bibfnamefont {J.~S.}\ \bibnamefont {Hofmann}}, \bibinfo {author}
  {\bibfnamefont {E.}~\bibnamefont {Huffman}}, \bibinfo {author} {\bibfnamefont
  {Z.}~\bibnamefont {Liu}}, \bibinfo {author} {\bibfnamefont {F.~P.}\
  \bibnamefont {Toldin}}, \bibinfo {author} {\bibfnamefont {J.~S.~E.}\
  \bibnamefont {Portela}},\ and\ \bibinfo {author} {\bibfnamefont
  {J.}~\bibnamefont {Schwab}},\ }\bibfield  {title} {\bibinfo {title} {{The ALF
  (Algorithms for Lattice Fermions) project release 2.0. Documentation for the
  auxiliary-field quantum Monte Carlo code}},\ }\href
  {https://doi.org/10.21468/SciPostPhysCodeb.1} {\bibfield  {journal} {\bibinfo
   {journal} {SciPost Phys. Codebases}\ ,\ \bibinfo {pages} {1}} (\bibinfo
  {year} {2022})}\BibitemShut {NoStop}%
\bibitem [{\citenamefont {Blankenbecler}\ \emph {et~al.}(1981)\citenamefont
  {Blankenbecler}, \citenamefont {Scalapino},\ and\ \citenamefont
  {Sugar}}]{Blankenbecler81}%
  \BibitemOpen
  \bibfield  {author} {\bibinfo {author} {\bibfnamefont {R.}~\bibnamefont
  {Blankenbecler}}, \bibinfo {author} {\bibfnamefont {D.~J.}\ \bibnamefont
  {Scalapino}},\ and\ \bibinfo {author} {\bibfnamefont {R.~L.}\ \bibnamefont
  {Sugar}},\ }\bibfield  {title} {\bibinfo {title} {Monte carlo calculations of
  coupled boson-fermion systems.},\ }\href
  {https://doi.org/10.1103/PhysRevD.24.2278} {\bibfield  {journal} {\bibinfo
  {journal} {Phys. Rev. D}\ }\textbf {\bibinfo {volume} {24}},\ \bibinfo
  {pages} {2278} (\bibinfo {year} {1981})}\BibitemShut {NoStop}%
\bibitem [{\citenamefont {White}\ \emph {et~al.}(1989)\citenamefont {White},
  \citenamefont {Scalapino}, \citenamefont {Sugar}, \citenamefont {Loh},
  \citenamefont {Gubernatis},\ and\ \citenamefont {Scalettar}}]{White89}%
  \BibitemOpen
  \bibfield  {author} {\bibinfo {author} {\bibfnamefont {S.}~\bibnamefont
  {White}}, \bibinfo {author} {\bibfnamefont {D.}~\bibnamefont {Scalapino}},
  \bibinfo {author} {\bibfnamefont {R.}~\bibnamefont {Sugar}}, \bibinfo
  {author} {\bibfnamefont {E.}~\bibnamefont {Loh}}, \bibinfo {author}
  {\bibfnamefont {J.}~\bibnamefont {Gubernatis}},\ and\ \bibinfo {author}
  {\bibfnamefont {R.}~\bibnamefont {Scalettar}},\ }\bibfield  {title} {\bibinfo
  {title} {Numerical study of the two-dimensional hubbard model},\ }\href
  {https://doi.org/10.1103/PhysRevB.40.506} {\bibfield  {journal} {\bibinfo
  {journal} {Phys. Rev. B}\ }\textbf {\bibinfo {volume} {40}},\ \bibinfo
  {pages} {506} (\bibinfo {year} {1989})}\BibitemShut {NoStop}%
\bibitem [{\citenamefont {Assaad}\ and\ \citenamefont
  {Evertz}(2008)}]{Assaad08_rev}%
  \BibitemOpen
  \bibfield  {author} {\bibinfo {author} {\bibfnamefont {F.}~\bibnamefont
  {Assaad}}\ and\ \bibinfo {author} {\bibfnamefont {H.}~\bibnamefont
  {Evertz}},\ }\bibfield  {title} {\bibinfo {title} {World-line and
  determinantal quantum monte carlo methods for spins, phonons and electrons},\
  }in\ \href {https://doi.org/10.1007/978-3-540-74686-7_10} {\emph {\bibinfo
  {booktitle} {Computational Many-Particle Physics}}},\ \bibinfo {series}
  {Lecture Notes in Physics}, Vol.\ \bibinfo {volume} {739},\ \bibinfo {editor}
  {edited by\ \bibinfo {editor} {\bibfnamefont {H.}~\bibnamefont {Fehske}},
  \bibinfo {editor} {\bibfnamefont {R.}~\bibnamefont {Schneider}},\ and\
  \bibinfo {editor} {\bibfnamefont {A.}~\bibnamefont {Wei{\ss}e}}}\ (\bibinfo
  {publisher} {Springer},\ \bibinfo {address} {Berlin Heidelberg},\ \bibinfo
  {year} {2008})\ pp.\ \bibinfo {pages} {277--356}\BibitemShut {NoStop}%
\bibitem [{Note1()}]{Note1}%
  \BibitemOpen
  \bibinfo {note} {As in Ref.~\cite {Goetz23}, we symmetrized the histograms by
  exploiting the $C_4$ symmetry of the model and the arbitrariness of the minus
  sign in the definition of the order parameter in Eq.~(\ref
  {eq:dimer_order}).}\BibitemShut {Stop}%
\bibitem [{\citenamefont {{Beach}}(2004)}]{Beach04a}%
  \BibitemOpen
  \bibfield  {author} {\bibinfo {author} {\bibfnamefont {K.~S.~D.}\
  \bibnamefont {{Beach}}},\ }\bibfield  {title} {\bibinfo {title} {{Identifying
  the maximum entropy method as a special limit of stochastic analytic
  continuation}},\ }\href {https://doi.org/10.48550/arXiv.cond-mat/0403055}
  {\bibfield  {journal} {\bibinfo  {journal} {arXiv e-prints}\ ,\ \bibinfo
  {eid} {cond-mat/0403055}} (\bibinfo {year} {2004})},\ \Eprint
  {https://arxiv.org/abs/cond-mat/0403055} {arXiv:cond-mat/0403055
  [cond-mat.str-el]} \BibitemShut {NoStop}%
\bibitem [{\citenamefont {Sandvik}(1998)}]{Sandvik98}%
  \BibitemOpen
  \bibfield  {author} {\bibinfo {author} {\bibfnamefont {A.}~\bibnamefont
  {Sandvik}},\ }\bibfield  {title} {\bibinfo {title} {Stochastic method for
  analytic continuation of quantum monte carlo data},\ }\href
  {https://doi.org/10.1103/PhysRevB.57.10287} {\bibfield  {journal} {\bibinfo
  {journal} {Phys. Rev. B}\ }\textbf {\bibinfo {volume} {57}},\ \bibinfo
  {pages} {10287} (\bibinfo {year} {1998})}\BibitemShut {NoStop}%
\bibitem [{\citenamefont {Shao}\ and\ \citenamefont {Sandvik}(2023)}]{Shao23}%
  \BibitemOpen
  \bibfield  {author} {\bibinfo {author} {\bibfnamefont {H.}~\bibnamefont
  {Shao}}\ and\ \bibinfo {author} {\bibfnamefont {A.~W.}\ \bibnamefont
  {Sandvik}},\ }\bibfield  {title} {\bibinfo {title} {Progress on stochastic
  analytic continuation of quantum monte carlo data},\ }\href
  {https://doi.org/https://doi.org/10.1016/j.physrep.2022.11.002} {\bibfield
  {journal} {\bibinfo  {journal} {Physics Reports}\ }\textbf {\bibinfo {volume}
  {1003}},\ \bibinfo {pages} {1} (\bibinfo {year} {2023})},\ \bibinfo {note}
  {progress on stochastic analytic continuation of quantum Monte Carlo
  data}\BibitemShut {NoStop}%
\bibitem [{\citenamefont {Assaad}\ and\ \citenamefont
  {Grover}(2016)}]{Assaad16}%
  \BibitemOpen
  \bibfield  {author} {\bibinfo {author} {\bibfnamefont {F.~F.}\ \bibnamefont
  {Assaad}}\ and\ \bibinfo {author} {\bibfnamefont {T.}~\bibnamefont
  {Grover}},\ }\bibfield  {title} {\bibinfo {title} {Simple fermionic model of
  deconfined phases and phase transitions},\ }\href
  {https://doi.org/10.1103/PhysRevX.6.041049} {\bibfield  {journal} {\bibinfo
  {journal} {Phys. Rev. X}\ }\textbf {\bibinfo {volume} {6}},\ \bibinfo {pages}
  {041049} (\bibinfo {year} {2016})}\BibitemShut {NoStop}%
\bibitem [{\citenamefont {Nandkishore}\ \emph {et~al.}(2012)\citenamefont
  {Nandkishore}, \citenamefont {Metlitski},\ and\ \citenamefont
  {Senthil}}]{Nandkishore12}%
  \BibitemOpen
  \bibfield  {author} {\bibinfo {author} {\bibfnamefont {R.}~\bibnamefont
  {Nandkishore}}, \bibinfo {author} {\bibfnamefont {M.~A.}\ \bibnamefont
  {Metlitski}},\ and\ \bibinfo {author} {\bibfnamefont {T.}~\bibnamefont
  {Senthil}},\ }\bibfield  {title} {\bibinfo {title} {Orthogonal metals: The
  simplest non-fermi liquids},\ }\href
  {https://doi.org/10.1103/PhysRevB.86.045128} {\bibfield  {journal} {\bibinfo
  {journal} {Phys. Rev. B}\ }\textbf {\bibinfo {volume} {86}},\ \bibinfo
  {pages} {045128} (\bibinfo {year} {2012})}\BibitemShut {NoStop}%
\bibitem [{\citenamefont {Hohenadler}\ and\ \citenamefont
  {Assaad}(2018)}]{Hohenadler18}%
  \BibitemOpen
  \bibfield  {author} {\bibinfo {author} {\bibfnamefont {M.}~\bibnamefont
  {Hohenadler}}\ and\ \bibinfo {author} {\bibfnamefont {F.~F.}\ \bibnamefont
  {Assaad}},\ }\bibfield  {title} {\bibinfo {title} {{Fractionalized Metal in a
  Falicov-Kimball Model}},\ }\href
  {https://doi.org/10.1103/PhysRevLett.121.086601} {\bibfield  {journal}
  {\bibinfo  {journal} {Phys. Rev. Lett.}\ }\textbf {\bibinfo {volume} {121}},\
  \bibinfo {pages} {086601} (\bibinfo {year} {2018})}\BibitemShut {NoStop}%
\bibitem [{\citenamefont {Hohenadler}\ and\ \citenamefont
  {Assaad}(2019)}]{Hohenadler19}%
  \BibitemOpen
  \bibfield  {author} {\bibinfo {author} {\bibfnamefont {M.}~\bibnamefont
  {Hohenadler}}\ and\ \bibinfo {author} {\bibfnamefont {F.~F.}\ \bibnamefont
  {Assaad}},\ }\bibfield  {title} {\bibinfo {title} {{Orthogonal metal in the
  Hubbard model with liberated slave spins}},\ }\href
  {https://doi.org/10.1103/PhysRevB.100.125133} {\bibfield  {journal} {\bibinfo
   {journal} {Phys. Rev. B}\ }\textbf {\bibinfo {volume} {100}},\ \bibinfo
  {pages} {125133} (\bibinfo {year} {2019})}\BibitemShut {NoStop}%
\bibitem [{\citenamefont {Xu}\ \emph {et~al.}(2019)\citenamefont {Xu},
  \citenamefont {Qi}, \citenamefont {Zhang}, \citenamefont {Assaad},
  \citenamefont {Xu},\ and\ \citenamefont {Meng}}]{Xu18}%
  \BibitemOpen
  \bibfield  {author} {\bibinfo {author} {\bibfnamefont {X.~Y.}\ \bibnamefont
  {Xu}}, \bibinfo {author} {\bibfnamefont {Y.}~\bibnamefont {Qi}}, \bibinfo
  {author} {\bibfnamefont {L.}~\bibnamefont {Zhang}}, \bibinfo {author}
  {\bibfnamefont {F.~F.}\ \bibnamefont {Assaad}}, \bibinfo {author}
  {\bibfnamefont {C.}~\bibnamefont {Xu}},\ and\ \bibinfo {author}
  {\bibfnamefont {Z.~Y.}\ \bibnamefont {Meng}},\ }\bibfield  {title} {\bibinfo
  {title} {{Monte Carlo Study of Lattice Compact Quantum Electrodynamics with
  Fermionic Matter: The Parent State of Quantum Phases}},\ }\href
  {https://doi.org/10.1103/PhysRevX.9.021022} {\bibfield  {journal} {\bibinfo
  {journal} {Phys. Rev. X}\ }\textbf {\bibinfo {volume} {9}},\ \bibinfo {pages}
  {021022} (\bibinfo {year} {2019})}\BibitemShut {NoStop}%
\bibitem [{\citenamefont {Shao}\ \emph {et~al.}(2016)\citenamefont {Shao},
  \citenamefont {Guo},\ and\ \citenamefont {Sandvik}}]{Shao15}%
  \BibitemOpen
  \bibfield  {author} {\bibinfo {author} {\bibfnamefont {H.}~\bibnamefont
  {Shao}}, \bibinfo {author} {\bibfnamefont {W.}~\bibnamefont {Guo}},\ and\
  \bibinfo {author} {\bibfnamefont {A.~W.}\ \bibnamefont {Sandvik}},\
  }\bibfield  {title} {\bibinfo {title} {Quantum criticality with two length
  scales},\ }\href {https://doi.org/10.1126/science.aad5007} {\bibfield
  {journal} {\bibinfo  {journal} {Science}\ }\textbf {\bibinfo {volume}
  {352}},\ \bibinfo {pages} {213} (\bibinfo {year} {2016})}\BibitemShut
  {NoStop}%
\end{thebibliography}%

\clearpage

\noindent
{\Large Supplemental Material for: Tuning the order of a deconfined quantum critical point}

% Includes S1, S2, ... in figures and Eqs
\makeatletter
\renewcommand{\thefigure}{S\@arabic\c@figure}
\renewcommand{\tablename}{Supplementary Table}
\renewcommand{\theequation}{S\@arabic\c@equation}
\setcounter{equation}{0}
\setcounter{figure}{0}
\makeatletter

\subsection*{Relevance of interaction range for the SO(4) model}

\begin{figure}[t]
  \includegraphics[width=1\linewidth]{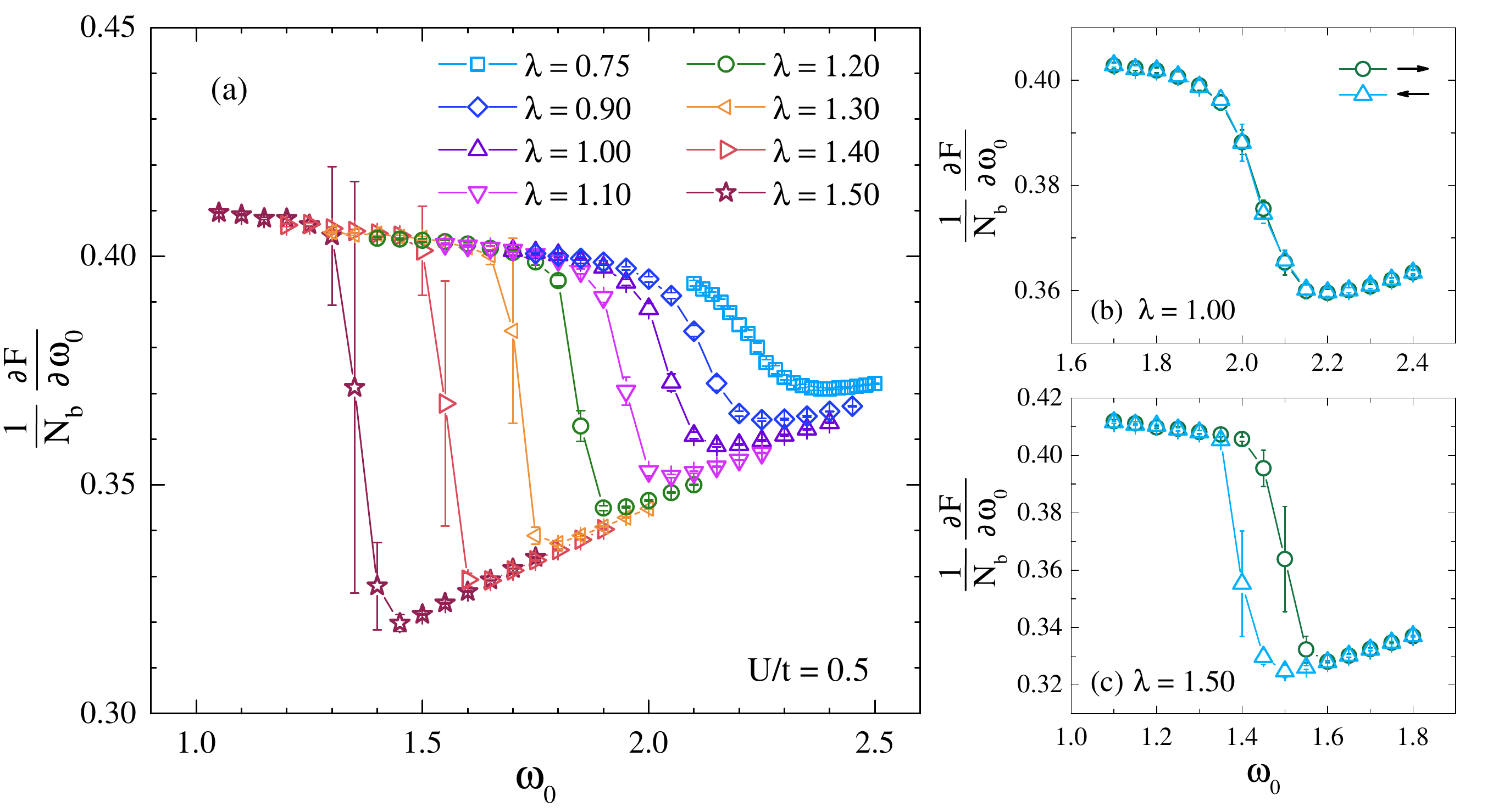}
   \caption{\label{fig:SSHH3} (a) Normalized free-energy derivative with respect to $\omega_{0}$. (b) and (c) Hysteresis curve for the free energy derivative with respect to $\omega_0$.  Here, we fixed $U=0.5$, $L =10$, $\beta=L$, and $t=0.1$.}
 \end{figure}

\begin{figure}[t]
 \includegraphics[width=1\linewidth]{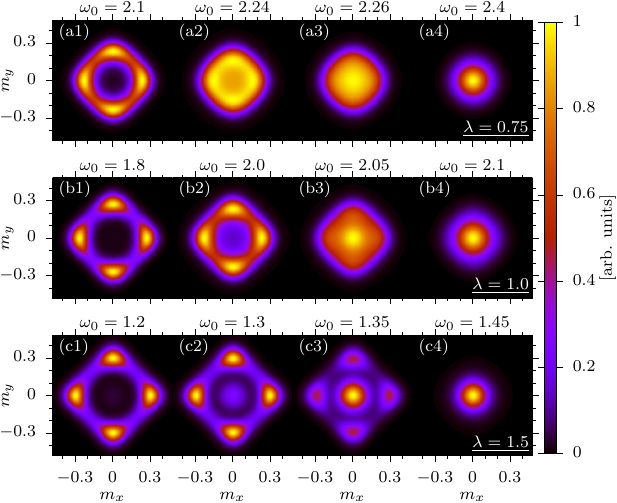}
  \caption{\label{fig:hist_SU2}Histogram of the VBS order parameter 
    $m_x$ and $m_y$  for different $\lambda$ and $\omega_0$ at $t=0.1$, $U=0.5$,
    $\beta=L=10$.  }
\end{figure}

Figures~\ref{fig:SSHH3} and \ref{fig:hist_SU2} depict the very same quantities as in Figs.~\ref{fig:scal_hyst_O4} and \ref{fig:hist_O4} but at $U=0.5$. Similar phenomena can be observed: upon increasing $\lambda$  and thereby decreasing the critical phonon frequency, a strong first-order transition emerges.

\end{document}